\documentclass[fleqn,twoside,twocolumn,nofootinbib]{revtex4} 
\usepackage{ujp} 

\begin{document}
\title[BINDING PARAMETERS OF ALKALOIDS BERBERINE]
{BINDING PARAMETERS OF ALKALOIDS BERBERINE AND SANGUINARINE WITH DNA}%
\author{V.G. Gumenyuk}
\affiliation{Taras Shevchenko National University of Kyiv}
\address{6, Academician Glushkov Prosp., Kyiv 03127, Ukraine}
\email{lns@univ.kiev.ua}
\author{N.V. Bashmakova}
\affiliation{Taras Shevchenko National University of Kyiv}
\address{6, Academician Glushkov Prosp., Kyiv 03127, Ukraine}
\email{lns@univ.kiev.ua}
\author{S.Yu. Kutovyy}
\affiliation{Taras Shevchenko National University of Kyiv}
\address{6, Academician Glushkov Prosp., Kyiv 03127, Ukraine}
\email{lns@univ.kiev.ua}
\author{V.M.~Yashchuk}
\affiliation{Taras Shevchenko National University of Kyiv}
\address{6, Academician Glushkov Prosp., Kyiv 03127, Ukraine}
\email{lns@univ.kiev.ua}
\author{L.A.~Zaika}
\affiliation{Institute of Molecular Biology and Genetics, Nat.
Acad. of Sci. of Ukraine}
\address{150, Academician Zabolotnyi Str., Kyiv 03143, Ukraine}
\udk{535.34; 535.37}  \razd{}

\setcounter{page}{524}%
\maketitle

\begin{abstract}
We study the interaction of berberine and sanguinarine (plant alkaloids) with DNA
in aqueous solutions, by using optical spectroscopy
methods (absorption and fluorescence). The dependencies of alkaloid
spectral characteristics on the concentration ratio $N/c$ between
the DNA base pairs and alkaloid molecules in the solutions are
considered, and the manifestations of the alkaloid--DNA binding are
revealed. The character of binding is found to depend on $N/c$. The
parameters of the binding of berberine and sanguinarine with DNA are
determined, by using the modified Scatchard and McGhee--von Hippel
equations.
\end{abstract}

\section{Introduction}

The creation of effective low-toxic antineoplastic preparations on the basis of
natural alkaloids is an important problem of modern medicine. Alkaloids are
actual for these preparations because of their property to be selectively
accumulated in tumor cells and their capability to form non-covalent complexes
with nucleic acids, by blocking the processes of transcription and replication of
the latter.

\begin{figure}[b]
\includegraphics[width=\column]{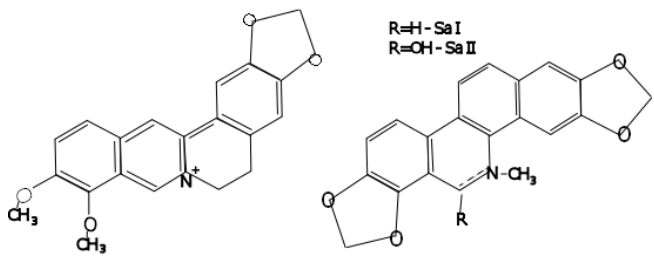}
\caption{Formula structures of berberine ($a$) and
sanguinarine ($b$). The dashed line marks a double bond that exists
in the SaI form}
\end{figure}

This work continues our researches [1] aimed at studying the interaction
between deoxyribonucleic acid (DNA) and celandine alkaloids, berberine and
sanguinarine. The latter two are included into the content of the
antineoplastic preparation, amitozine, created at the Institute of Molecular
Biology and Genetics of the National Academy of Sciences of Ukraine. Berberine
and sanguinarine (Fig.~1) belong to the isoquinoline group. The structural
formulas of berberine and sanguinarine molecules are C$_{20}$H$_{19}$NO$_{5}$
and C$_{20}$H$_{15}$NO$_{5}$, respectively.

Usually, isoquinoline alkaloids interact with DNA as intercalators, or they
are arranged in a small groove; their external binding with phosphate groups
is also possible. However, the binding mechanisms have not been definitively
determined---neither for berberine, nor for sanguinarine---despite that rather
intensive researches of this issue have been carried out.

The interaction between berberine and nucleic acids was studied using spectral
methods in a number of works, but conclusions concerning the way of their
binding are ambiguous. Some results testify to the intercalation (complete
[2] or partial [3-5]), whereas the others to groove binding
[6,7]. In work [8], two independent types of binding with different
affinities were proposed, and the association constant was found for each of them.

The interaction between sanguinarine and DNA was also studied in a
number of works (see review [9]). Sanguinarine is known
[9] to exist in aqueous solutions in two forms, imine (SaI,
$\mathrm{pH}<6$) and alkanolamine (SaII, $\mathrm{pH}>8.5$). The
former is a cation, the latter is neutral (see Fig.~1). In work
[10], the formation of complexes partially intercalated into
DNA was shown for both sanguinarine forms, imine and alkanolamine.
However, this conclusion was denied by the results of work
[11], the authors of which asserted that only one form, imine,
interacts with DNA. At the same time, the intercalation as a way for
sanguinarine to bind with DNA was confirmed in work [12].

In this work, we obtained the dependencies
of the spectral characteristics of berberine and sanguinarine on the ratio $N/c$ between the number of DNA base
pairs and the number of alkaloid molecules. We also specified
manifestations of the binding between those alkaloids and DNA. A special attention
was given to the analysis of the Scatchard and McGhee--von Hippel equations and
their application to the numerical approximation of experimental data.

\section{Experimental Specimens and Technique}

We used alkaloids berberine (Be, \textquotedblleft Alps
Pharmaceutical\textquotedblright, Japan) and sanguinarine (Sa, Ivan
Franko Lviv National University) fabricated in the form of
microcrystalline powders. The latter were dissolved in water for
injections at a temperature within the interval of
60--70$~^{\circ}\mathrm{C}$. The experimental alkaloid
concentrations ranged from 12.5 to 50~$\mu$M. Reabsorption or
concentration effects are insignificant at such concentrations. We
used the DNA of chicken erythrocytes (DNA CE) treated with
ultrasound and the DNA of calf thymus (DNA CT), both were obtained
from Serva (Heidelberg, Germany). The average molar mass of a
nucleotide pair was about 650~Da. When measuring the concentration
dependences for the solutions Be + DNA and Sa + DNA, the
concentration of an alkaloid remained constant, but the
concentration of DNA varied. The ratio between the molar
concentrations of DNA and an alkaloid ($N/c$) was expressed in terms
of the number of nucleotide pairs per one alkaloid molecule.

The absorption spectra were registered on a Specord UV~VIS
spectrophotometer in the range of 200--700\textrm{~nm}. The spectral
resolution was 1\textrm{~nm}. Fluorescence spectra were obtained making use of
a Cary Eclipse fluorometer\ in the range of 300--800\textrm{~nm}. The spectral
width of a slit for fluorescence measurements was 5\textrm{~nm}.

\section{Experimental Results}

The interaction of alkaloids with DNA is characterized by such phenomena in
absorption and fluorescence spectra as the hypochromism in alkaloid absorption
bands; the \textquotedblleft red\textquotedblright\ and \textquotedblleft
blue\textquotedblright\ shifts of maxima in the absorption and fluorescence,
respectively, spectra; and the variation of the fluorescence quantum yield. These
and other manifestations of the binding of berberine and sanguinarine with DNA
observed in optical spectra and their dependencies on $N/c$ will be analyzed in
detail elsewhere. Here, we only describe them in brief.

\begin{figure}
\includegraphics[width=\column]{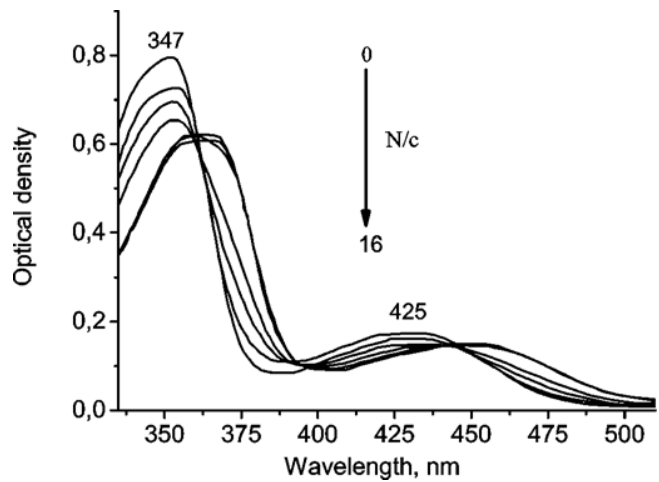}
\caption{Absorption spectra of the aqueous solution of
berberine and complex Be + DNA~CT at various values of $N/c$.
$c_{\mathrm{Be}}=2.5\times10^{-5}$~M  }
\end{figure}

\subsection{Berberine}

The absorption spectrum of berberine lies in the spectral interval
$\lambda<500$\textrm{~nm} and consists of four two-component bands with the
maxima at 425, 347, 264, and 229\textrm{~nm}. When DNA molecules are added to
the berberine solution, a substantial hypochromism (up to 30 \%) and the shift
of maxima toward long waves (up to 22\textrm{~nm}) are observed for 345 and
425-nm bands in the alkaloid absorption spectrum, which testifies to the
binding of berberine with DNA (Fig.~2).

The dipole moments of planar aromatic molecules is oriented in the molecular
plane [13]. Therefore, the spatial arrangement of the dipole moments of
those molecules corresponds to that of the molecules themselves. If the dipole
moments of molecules are oriented in parallel (the \textquotedblleft
sandwich\textquotedblright\ structure), the spectra demonstrate hypochromism.
In this case, hypochromism, as a consequence of the interaction between the $\pi
$-systems of molecules oriented in parallel to one another, testifies to
the intercalation or external stacking as probable binding mechanisms.

The fluorescence spectrum of berberine has one band with a maximum at about
556\textrm{~nm}. The fluorescence quantum yield of berberine is very low at
room temperature. The fluorescence spectra of the complex Be--DNA are
characterized by a very considerable amplification of their intensity (up to a
factor of 200, depending on the excitation $\lambda$, the concentration, and
the specimen type) and a blue shift of the fluorescence maximum (up to
26\textrm{~nm}) in comparison with the free berberine case (Figs.~3 and~4).

\begin{figure}
\includegraphics[width=\column]{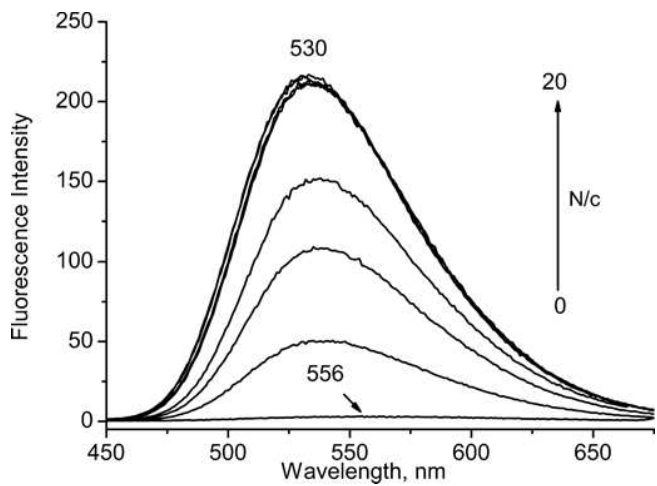}
\caption{Fluorescence spectra of berberine and complex Be
+ DNA~CT at various values of $N/c$.
$c_{\mathrm{Be}}=4.1\times10^{-5}$~M. Excitation at 450\textrm{~nm} }
\end{figure}

\begin{figure}
\includegraphics[width=\column]{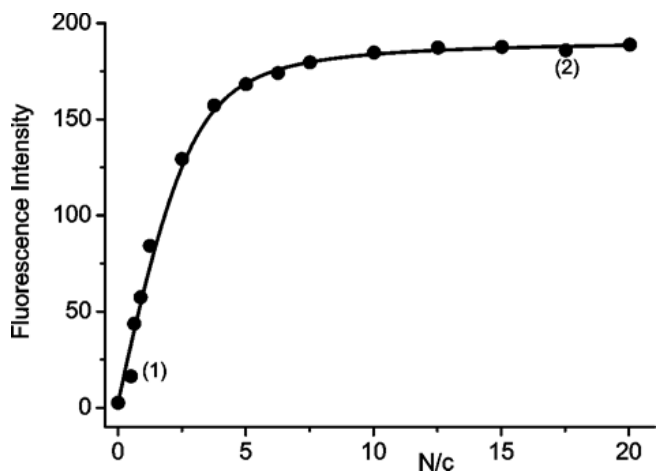}
\caption{Dependence of the fluorescence intensity of
complex Be + DNA~CT on $N/c$. $c_{\mathrm{Be}}=4.1\times10^{-5}$~M.
Excitation at 450\textrm{~nm}. As for points (1) and (2), see
section 4.5 }
\end{figure}

The changes in the fluorescence spectra are caused by the fixation of
berberine molecules on the DNA matrix. Namely, the probability of
the radiationless excitation relaxation diminishes at the complex formation, because
the interaction of vibrations with one another is less effective. Moreover, the
energy transfer to solvent molecules is also less effective now. Accordingly,
the fluorescence quantum yield grows. The \textquotedblleft
blue\textquotedblright\ shift of the fluorescence maximum is mainly associated
with variations in the polarity of the fluorophore molecule environment: DNA
is a less polar medium for alkaloid molecules than water. In addition, bound
alkaloid molecules are partially screened from solvent molecules. As a result,
the effect of solvent relaxation, i.e. the interference between the dipole
moments of excited alkaloid and solvent molecules [14] becomes insignificant.

Following the procedure of work [15], we found the intersection point of
the fluorescence band and the first absorption band plots presented in terms
of $I/\nu^{4}$ and $\varepsilon/\nu$ units, respectively, to determine the
dependence of the frequency of the first electron transition (0-0) in the
system Be + DNA on $N/c$.

In whole, all the mentioned dependencies (the \textquotedblleft
red\textquotedblright\ and \textquotedblleft blue\textquotedblright\ shifts,
variations of the fluorescence intensity and the hypochromism degree, a
change of the 0-0 transition frequency) are characterized by a mutual
behavior of the type depicted in Fig.~4: a rather quick growth of the
corresponding parameter at the beginning followed by a saturation at certain,
close to one another, $N/c$-values.

\begin{figure}
\includegraphics[width=\column]{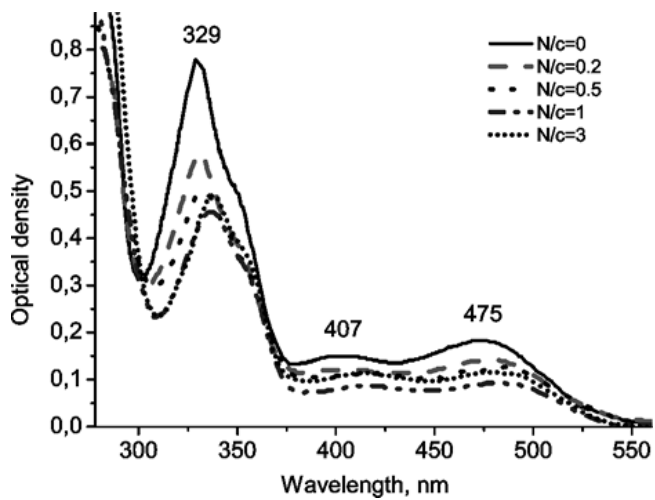}
\caption{Absorption spectra of a solution Sa + DNA~CE at
various values of $N/c$. $c_{\mathrm{Sa}}=5\times10^{-5}$~M}
\end{figure}

\subsection{Sanguinarine}

As was mentioned, the sanguinarine molecule can exist in the imine
(at $\mathrm{pH}<6$) or alkanolamine (at $\mathrm{pH}>8.5$) form. In
our experiments, pH was about 7, so that both forms were available.
The presence of two Sa forms in the solution is characterized by a
complicated shape of the absorption curve. The latter includes bands
typical of two Sa forms (Fig.~5). The fluorescence spectrum consists
of two bands with the maxima at 587 and 419\textrm{~nm} (Fig.~6),
which different excitation spectra correspond to.

\begin{figure}
\includegraphics[width=\column]{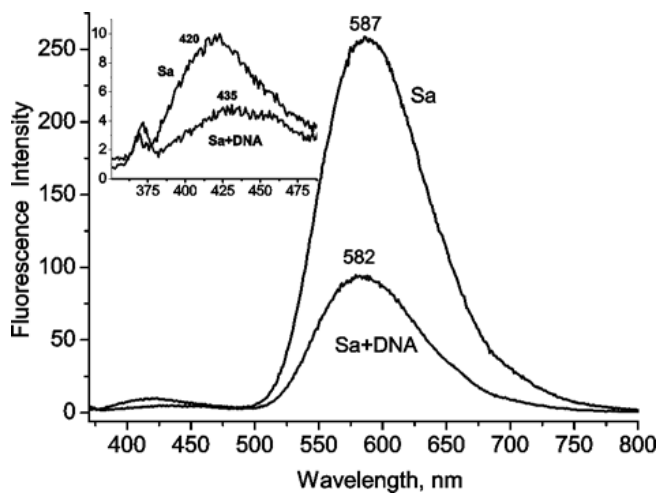}
\vskip-3mm
\caption{Fluorescence spectra of sanguinarine and complex
Sa + DNA~CT at $N/c=1$. $c_{\mathrm{Sa}}=5\times10^{-5}$~M. Excitation at 330\textrm{~nm}. The corresponding spectra of the SaII form are shown in the inset}
\end{figure}

When adding DNA, the absorption and fluorescence spectra of
sanguinarine, similarly to those of berberine, also demonstrate the
band shift, hypochromism, and a variation of the fluorescence
intensity. However, in contrast to berberine, variations of
the optical density and the fluorescence intensity, as well as the frequency
of the first electron transition (0-0), depend on the ratio $N/c$
for sanguinarine in a nonstandard way. Namely, the corresponding
curves have a minimum (see the fluorescence spectrum in Fig.~7), and
such a behavior is observed for both sanguinarine forms.

The minimum in the fluorescence intensity contradicts the data of work
[19]. A similar dependence was found only in work [16], but no
relevant explanation was given. This atypical variation of the optical parameter
of alkaloid can be explained as a manifestation of two types of the binding of sanguinarine
to DNA: an external one and the intercalation. The minima in the
dependencies at $N/c\leq1$ correspond to the most compact arrangement of
sanguinarine molecules on the DNA matrix, which brings about the hypochromism in
the absorption bands (by about 40--50\%) and the fluorescence quenching (by about
70--80\%). Since the intercalation mechanism of binding is characterized by
the values $N/c\geq2$, the most probable mechanism of binding at $N/c<1$ is
the mechanism of external \textquotedblleft stacking\textquotedblright. When
the DNA concentration grows, the number of binding sites also increases, and
alkaloid molecules can be arranged at a larger distance from one another along
the DNA chain, which is observed as an increase of the optical parameter.
In this case, the most probable for sanguinarine is the
intercalation way of binding. At certain concentration ratios, the dependencies saturate.

\begin{figure}
\includegraphics[width=8.3cm]{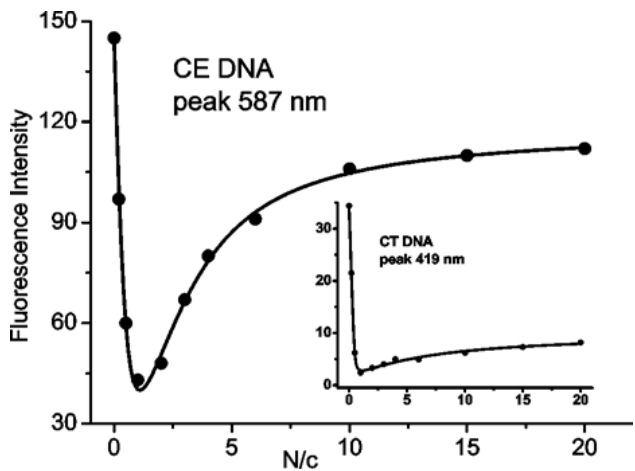}
\caption{Dependence of the fluorescence intensity on $N/c$
for the 587-nm band of complex Sa + DNA~CE. Excitation at
470\textrm{~nm}. The same but for the 419-nm band and the excitation
at 330\textrm{~nm} is shown in the inset.
$c_{\mathrm{Sa}}=5\times10^{-5}$~M  }
\end{figure}

Note that, although the character of the intensity variation is similar for the
fluorescence bands at 419 and 587\textrm{~nm}, the minimum for the former band
is observed at larger $N/c$-values. This fact evidences the interaction
between SaII and DNA, which is more effective at larger $N/c$-ratio values in
comparison with that for the SaI form.

As the main reason of the fluorescence yield reduction in the case of
sanguinarine, we consider an effect similar to the concentration quenching. In
particular, by binding with DNA (it is especially valid for the external
binding), alkaloid molecules are arranged relatively close to one another in
comparison with free molecules, which is responsible for some fluorescence quenching.

We also note that all the mentioned dependencies for the DNA~CE and DNA~CT have
the same character, but, in the DNA~CT case, they attain the corresponding
maximum or minimum (or saturate) at smaller $N/c$-values. It looks as if the
alkaloid interacts more effectively with DNA~CT, which was no subjected to
the ultrasonic treatment. The corresponding calculations for binding parameters
(see below) confirmed this assumption.

\section{Calculation of Binding Parameters}

\subsection{Theory}

To determine the parameters of small ligand binding with DNA, we wrote a
computer program BindFit. Its feature is the operation with direct
experimental data, i.e. the absence of any coordinate transformations known as
\textquotedblleft linearization\textquotedblright. It allowed us to improve
the accuracy of the parameter determination for processes described by nonlinear
plots even after the linearization (see section 4.5).

A classical equation for the determination of binding parameters (more
specifically, the association constant) is the Scatchard equation [17],
\begin{equation}
\frac{\nu}{c_{f}}=K(1-\nu)
\end{equation}
Here, $\nu$ is the ratio between the concentration of bound ligands $c_{b}$
and the total concentration of binding sites $N$, $c_{f}$ is the concentration
of free ligands, and $K$ is the association constant. According to
Eq.~(1), the dependence of the ratio $\nu/c_{f}$ on $\nu$ must be
linear, and the association constant can be determined as the slope of the
obtained straight line. For some reasons, this dependence can be nonlinear, in
particular, if a ligand molecule occupies more than one binding site (namely,
$n$ sites) in the DNA matrix. Then, the equation looks like $\nu
/c_{f}=K(1-n\nu)$. However, its use is not always correct. Owing to the
underestimation of the number of empty binding sites, the binding parameters are
determined incorrectly. In particular, the association constant can
be overestimated by a factor of about $2n$ [18].

McGhee and von Hippel [18] developed Scatchard's approach, by
extending it onto the case $n>1$. Taking advantage of the
probability theory methods, they correctly made allowance for that
fact that extended ligand chains can occupy more than one binding
site. The McGhee--von Hippel equations are as follows:

\noindent--~for the non-cooperative binding,
\begin{equation}
\frac{\nu}{c_{f}}=K(1-n\nu)\left( \frac{1-n\nu}{1-(n-1)\nu}\right)
^{n-1}
\end{equation}

\noindent--~for the cooperative binding,
\[
 \frac{\nu}{c_{f}}=K(1-\nu)
 \left(
\frac{(2\omega-1)(1-n\nu)+\nu-R}{2(\omega-1)(1-n\nu)}\right)^{n-1}
\times
\]
\begin{equation}
\times \left(\frac{1-(n+1)\nu+R}{2(1-n\nu)}
 \right)^{2}.
\end{equation}

\noindent Here,
\[
R=\sqrt{(1-(n+1)\nu)^{2}+4\omega\nu(1-n\nu)},
\]
and $\omega$ is the parameter of cooperativity ($\omega>1$ for cooperative,
$\omega=1$ for non-cooperative, and $\omega<1$ for anti-cooperative binding).

In most cases, the plots are nonlinear in the Scatchard coordinates
$(\nu,\nu/c_{f})$. However, in terms of those variables, the binding equation can
be written down in an explicit form, which is impossible to be done in
experimental variables. That is why the Scatchard variables found a definite application.

\subsection{Connection with the optical parameter}

The concentrations of bound or free ligands are not experimentally observable
quantities. Experimentally, the dependencies of the optical parameters of a
solution on the concentration of its components are measured. Usually, the
optical parameter is calculated as a sum of contributions made by ligands in
both states, bound and free.

\subsubsection*{4.2.1. Two states of ligand (one type of binding sites)}

Provided that the ligand can be in only one of two states (free or bound), the
expression for the optical parameter is rather simple, and it allows the
concentrations of free and bound ligands to be determined straightforwardly:

\begin{equation}\label{4}
A=\frac{c_{f}}{c}A_{f}+\frac{c_{b}}{c}A_{b},%
\end{equation}
where $A$ is the optical parameter of the mixture, $c$ the total
concentration of ligands, $A_{f}$ the optical parameter for free
ligands, $c_{f}$ the free ligand concentration, $A_{b}$ the optical
parameter for bound ligands, $c_{b}$ the bound ligand concentration,
$c=c_{f}+c_{b}$. Knowing the parameters $A_{f}$ and $A_{b}$ (the
former is determined for the solution of a pure ligand, the latter
could be determined using the solution of a ligand with a
considerable excess of DNA), it is possible to determine $c_{f}$ and
$c_{b}$, and then change to the Scatchard coordinates:
\begin{equation}
\nu=\frac{(A-A_{f}) c}{(A_{b}-A_{f}) N},  \quad
\frac{\nu}{c_{f}}=\frac{A-A_{f}}{(A_{b}-A) N}.
\end{equation}
For the approximation of direct experimental data, those procedures
are redundant.

\subsubsection*{4.2.2. Three states of ligand (two types of binding sites)}

Provided that the ligand can be in only one of three states (free, bound to
a site of the first type, and bound to a site of the second type), the
expression for the optical parameter $A$ looks like
\begin{equation}
A=\frac{c_{f}}{c}A_{f}+\frac{c_{b}^{(1)}}{c}A_{b}^{(1)}+\frac{c_{b}^{(2)}}%
{c}A_{b}^{(2)},%
\end{equation}
where $A_{b}^{(1)}$ and $A_{b}^{(2)}$ are the optical parameters of ligands
bound to sites of the first and the second type, respectively; and
$c_{b}^{(1)}$ and $c_{b}^{(2)}$ are the corresponding concentrations of those
ligands. In this case, it is difficult to determine the concentration of
ligands in each state (and, hence, to change to the Scatchard coordinates),
because both $A_{b}^{(1)}$ and $A_{b}^{(2)}$ cannot be determined by direct
measurements, so that the number of equations in the system used for the
determination of the concentrations of bound and free ligands is smaller than the
number of variables. This problem does not arise at a straightforward
approximation of experimental data, because those quantities are calculated
together with other parameters.

\subsection{Model equations}

For numerical analysis, Eqs.~(1--3) written down in terms of
the $\nu$ and $\nu/c_{f}$ variables are not convenient, because those quantities
are connected with experimental optical parameters (the fluorescence
intensity, the optical density, and others) in rather a complicated way.
In addition, the equations for the binding parameters written down in terms of
the Scatchard variables give solutions with a relatively high error. It is
associated with the fact that, in the general case of nonlinear dependences,
the linearization \textquotedblleft following Scatchard\textquotedblright%
\ gives rise to a considerable distortion of experimental errors and,
respectively, the determination accuracy for binding parameters, which should
better be determined from the initial, non-linearized curves. Therefore, in
order to analyze and to directly (i.e. in terms of experimental variables)
approximate experimental data, the indicated equations were modified and
applied in the form of equations with the independent variable $c_{b}$ , i.e.
the concentration of bound ligands.

\subsubsection*{4.3.1. One type of binding sites}

In this case, basic are the McGhee--von Hippel equations transformed from
their original form to that including only the variables directly related to
the experiment. As a rule, what is experimentally measured is the dependence
of a certain optical parameter of the solution on the concentration ratio
between the dissolved components; normally, it is the ratio between the total
concentration of binding sites to the total concentration of ligands, $N/c$,
i.e. the quantity reciprocal to $\nu$. It is the concentrations of components
rather than those of bound and free ligands that are known. Therefore, such
variables as the total concentrations of ligands, $c$, and binding sites, $N$,
would be more expedient for computerized processing. In this case, there
remains only one unknown variable in the equation, which can be determined numerically.

Carrying out a series of simple transformations (we multiply the
\textquotedblleft canonical\textquotedblright\ equations (2)
or (3) by the factors $c_{f}$ and $N$, and make some changes
in the variable notations), the McGhee--von Hippel equations for the
non-cooperative (Eq.~(2)) and cooperative (Eq.~(3))
bindings are reduced to the following sought expressions, which
include the variable $c_{b}$:
\begin{equation}
K(c-c_{b})(N-nc_{b})\left(\frac{N-nc_{b}}{N-(n-1)c_{b}}
\right)^{n-1}-c_{b}=0,
\end{equation}
\[
 K(c-c_{b})(N-nc_{b})
\left(\frac{(2\omega-1)(N-nc_{b})+c_{b}-R'}{2(\omega-1) (N-nc_{b})}
\right)^{n-1} \times
\]
\begin{equation}
\times \left(\frac{N-(n+1)c_{b}+R'}{2(N-nc_{b})}
\right)^{2}-c_{b}=0,
\end{equation}
where
\[
(R'=\sqrt{(N-(n+1)c_{b})^{2}+4\omega c_{b}(N-nc_{b})} ).
\]
It is those equations which were implemented in the program and were
solved numerically (we did not manage to obtain their analytical
solutions). It enabled us to operate with the quantity $c_{b}$ as
with the function $c_{b}=c_{b}(N,c;K,n)$.

A separate remark should be made on the parameter $n$. In the case $0<n<1$,
the substitutions $n=1$ and $N^{\prime}=N/n$\ are made by force in the
equation. In other words, such parameter values are interpreted as a hint
that, actually, there are more binding sites than $N$. This situation can be
realized, e.g., if $N$ stands for the concentration of DNA base pairs, and
binding occurs with phosphate residues, the concentration of which is $2N$.
However, the aforementioned $n$-values can also testify to the cooperative
binding. In this case, the process is described by the cooperative equation (8)
with $n=1$, with the compulsory condition $\omega>1$. At $\omega\rightarrow1,$
the denominator in Eq.~(8) tends to zero. If the corresponding passage
to the limit is done, the equation transforms into a non-cooperative one. This
situation is handled programmatically by changing from the cooperative
equation to the non-cooperative one at $\omega=1$. The parameter $n$ is
processed in the same way, as for the non-cooperative equation.

\subsubsection*{4.3.2. Two types of binding sites}

Processes with two types of binding sites can be schematically designated as
$c_{b}^{(1)}\Longleftrightarrow c_{f}\Longleftrightarrow c_{b}^{(2)}$, i.e.
the process of direct transition by bound ligands from sites of type 1 to
sites of type 2 is impossible. These processes are described by a system of
two equations, which must take into account whether the processes of binding
of the ligands that occupy one binding site (i.e. the base pair and phosphates)
are interdependent or not. If the parameter $N$ stands for the concentration
of DNA base pairs, then $2N$ binding sites correspond to the first type of
binding (with a phosphate) and $N$ binding sites to the second type (intercalation).

Among the implemented schemes, the simplest is the system of two modified
Scatchard equations, which describes two independent processes of binding of the
ligands that occupy one binding site:
\begin{equation}
\left\{
\begin{array}{l}
c_{b}^{(1)}=K_{1}(c-c_{b}^{(1)}-c_{b}^{(2)})(2N-c_{b}^{(1)}),\\[2mm]
c_{b}^{(2)}=K_{2}(c-c_{b}^{(1)}-c_{b}^{(2)})(N-c_{b}^{(2)}).
\end{array}
\right.
\end{equation}
Other combinations of the Scatchard and McGhee--von Hippel equations are
also possible.

For our experimental data, the best approximation results were
obtained making use of a system of modified Scatchard (for external
binding) and McGhee--von Hippel (for intercalation) equations. The
system describes two interdependent processes of binding of the ligands
that occupy one binding site. The intercalation into the interval
between the base pairs is allowed only if both phosphates in this
interval are not connected with ligands and {\it vice versa}, i.e. the binding
with phosphates is possible only if no ligand has intercalated into
the corresponding interval. In addition, there can be not less than
$n-1$ free intervals between two intercalated ligands. This model
brings about the following system of equations:
\begin{equation}
\left\{
\begin{array}{l}
c_{b}^{(1)}=K_{1}(c\!-\!c_{b}^{(1)}\!-\!c_{b}^{(2)})(2N-c_{b}^{(1)})\left(1\!-\!\frac{c_{b}^{(2)}}{N}\right),\\[2mm]
c_{b}^{(2)}=K_{2}(c-c_{b}^{(1)}-c_{b}^{(2)})(N-nc_{b}^{(2)})
\times\\[2mm]
\times \left(\frac{N-nc_{b}^{(2)}}{N-(n-1)c_{b}^{(2)}} \right)^{n-1}
 \left(1-\frac{c_{b}^{(1)}}{2N} \right)^{2}.
\end{array} \right. 
\end{equation}

It is important that an additional multiplier emerges in the equations for
interdependent processes. The mechanism of its appearance is as follows. Any
equation describing the binding looks like $c_{b}=Kc_{f}N_{f}$, i.e. the
concentration of bound ligands is equal to the product of the free ligand
concentration, the concentration of empty binding sites, and the association
constant. It is the factor $N_{f}$ that generates additional multipliers,
because this quantity ultimately acquires the form $N_{f}=(N-nc_{b})P_{f}$,
i.e. the concentration of binding sites is equal to the concentration of
unoccupied sites times the probability that no factors that prohibit the binding
are actual for any arbitrarily selected empty binding site. For interdependent
processes, this factor is selected to be the presence of a bound ligand at
the neighbor binding site, provided that this ligand belongs to the different
type. The probability that this factor does not interfere the binding is equal to
the probability that all neighbor binding sites are free, i.e.
\[
P_{f}=\left(  1-\frac{c_{b}^{\rm other}}{N^{\rm other}}\right)
^{s},
\]
where $s$ is the number of neighbor binding sites. In this case, this quantity
is determined as follows:

\noindent--~for the binding with a phosphate, it is a fraction of phosphates
belonging to the unoccupied interval,%
\[
p_{a}=1-\frac{2c_{b}^{(2)}}{2N}=1-\frac{c_{b}^{(2)}}{N};
\]
i.e. every occupied interval forbids the binding with two phosphates;

\noindent--~for the intercalation, it is a probability that both phosphates in
this interval are free,
\[
p_{b}=(P_{\rm ph.free})^{2}=(1-P_{\rm ph.occupied})^{2}=(1-\frac{c_{b}^{(1)}}{2N}%
)^{2};
\]
i.e. the probability that the phosphate is busy, $P_{\mathrm{ph.occupied}}$,
is equal to the probability that one of $c_{b}^{(1)}$-ligands became bound
with one of $2N$ phosphates.

Note that

\noindent1.~The processes, in which a direct transition of bound ligands from
the sites of type 1 to sites of type 2 is possible, is not considered here,
because several equations and about a dozen parameters are needed for their
description. It is too much for the parameters to be determined with a
satisfactory accuracy.

\noindent2.~Since the solution describes the concentration of bound ligands,
certain restrictions are imposed on it: the concentration of bound ligands
must be a non-negative number, it cannot exceed the total ligand
concentration, and it is confined from above by the concentration of binding sites.

Generally speaking, the equations given above have a number of solutions.
However, it turned out that only one of them satisfies the indicated
restrictions in the working range of parameters.

\noindent3.~Among the algorithms applied to solve the equations, the method of
binary search turned out to be the best, i.e. it produced stable results at
the highest calculation rate. Every parameter was approximated separately.

\subsection{Parameters of the complex formation for berberine and sanguinarine}

To determine the binding parameters, the most convenient is to use
the data on variations of both the optical density in the absorption
spectra and the fluorescence intensity. In our
experiments, more exact data were obtained from the fluorescence
spectra; accordingly, results of the corresponding approximation
turned out a little more accurate.

\subsubsection*{4.4.1. Berberine}

For the approximation of experimental data, we used the modified
McGhee--von Hippel equations (7) and (8) for the
non-cooperative and cooperative, respectively, bindings. It turned
out that the best results were obtained, if the binding was
considered to be cooperative, but the degree of cooperativity  was
small. For DNA~CT, we obtained the values quoted in Table~1. The
table demonstrates a very good agreement between the parameters
determined by approximating the data obtained in independent
experiments (fluorescence and absorption). The value $n\approx2$
means that one alkaloid molecule occupies two DNA base pairs, which
evidences the intercalation model of berberine binding to DNA. A
small cooperativity can testify to a certain untwisting of the DNA
helix at intercalation sites.


\begin{table}[b]
\noindent\caption{Parameters of the binding of berberine with
DNA}\vskip3mm\tabcolsep8.1pt

\noindent{\footnotesize\begin{tabular}{c c c}
 \hline \multicolumn{1}{c}
{\rule{0pt}{9pt}} & \multicolumn{1}{|c}{From fluorescence spectra}&
\multicolumn{1}{|c}{From absorption spectra}\\%
\hline%
$K$&($5.2\pm 0.2)\times10^{4}$&$(5.7\pm 0.7)\times 10^{4}$\\
$n$&$1.85\pm 0.1$&$1.8\pm0.3$\\%
$\omega$&$1.3\pm0.2$&$1.45\pm0.4$\\%
\hline
\end{tabular}}
\end{table}

For the sake of comparison, note that the values
$K=3.54\times10^{4}$ and $n=2$ were obtained in work [19] on the basis of
analysis of other equations (but in the Scatchard coordinates). Hence, we have good agreement
with those results, taking into account the difficulties faced when obtaining
exact experimental and calculation data.

\subsubsection*{4.4.2. Sanguinarine}

Since the dependencies of the optical density and the fluorescence
intensity on the DNA concentration in the solutions Sa + DNA are
nontrivial (their behavior is similar to what is shown in Fig.~7),
the McGhee--von Hippel equation cannot be applied directly to this
case. As turned out, the relevant experimental results are better
described by a system of equations, which takes two interdependent
processes of ligand binding into account; these are the external
binding with phosphates (type~1) and the intercalation into the DNA
double helix (type~2). In addition, the process of direct transition
by bound ligands from sites of type~1 onto sites of type~2 is
impossible (condition~1), and there must be not less than $n-1$ free
intervals between two intercalated ligands (condition~2). This model
gives rise to the system of equations (10), in which
conditions 1 and 2 are taken into account. The number of base pairs
occupied by one alkaloid at the external binding was considered to
be equal to 0.5; at the intercalation, $n$ was one of the equation
parameters. The concentration of external binding sites was adopted
to be about $2N$ and that of intercalation sites to be $N$.

By approximating the experimental dependencies with the systems of
equations (10), we obtained the values for the association
constant (in terms of M$^{-1}$ units) and the parameter $n$, which
are presented in Table~2.

\subsection{Characteristic features of Scatchard coordinates}

\label{sec4.5}The usage of the Scatchard coordinates requires that the
concentrations of bound and free ligands be determined directly from
experimental data, which is not always possible. However, even if it is
possible, this operation can distort the results uncontrollably. We intend to
illustrate the expediency of the used programmatic approach on a specific
example. For the comparison between the results obtained (namely, the $K$- and
$n$-values) to be more correct, we used the McGhee--von Hippel equation for
the non-cooperative binding, because the parameter $\omega$ cannot be determined
at all from the Scatchard equation.

\begin{table}[b]
\noindent\caption{Parameters of the binding of sanguinarine with
DNA}\vskip3mm\tabcolsep5.6pt

\noindent{\footnotesize\begin{tabular}{c c c c }
 \hline \multicolumn{1}{c}
{\rule{0pt}{9pt}} & \multicolumn{1}{|c}{SaI + DNA CE}&
\multicolumn{1}{|c}{SaI + DNA CT}&
\multicolumn{1}{|c}{SaII + DNA CT}\\%
\hline%
$K_{1}$&$(6.7\pm0.4)\times 10^{4}$&$(3.5\pm 0.3)\times10^{5}$&$(7.9\pm 0.3)\times10^{5}$\\
$K_{2}$&$(5.5\pm 0.7)\times 10^{5}$&$(1.1 \pm 0.4)\times10^{6}$&$(1.0\pm 0.5)\times10^{5}$\\%
$n$&$3.4\pm 0.3$&$1.5\pm 0.4$&$12\pm 3$\\%
\hline
\end{tabular}}
\end{table}

In Fig.~4, the raw experimental data for the dependence of the
fluorescence intensity on the ratio $N/c$ obtained for the complex
Be + DNA~CT and the results of their direct approximation are shown.
The approximation curves calculated using the equations for both the
cooperative (Fig.~4) and non-cooperative bindings practically
coincide. The differences become noticeable between the numerical
values of the parameters determined making use of the corresponding
equations. For the non-cooperative binding, we obtained
$K=(7.69\pm3.18)\times10^{4}~\mathrm{M}^{-1}$ and $n=2.11\pm0.23$.

\begin{figure}
\includegraphics[width=8.3cm]{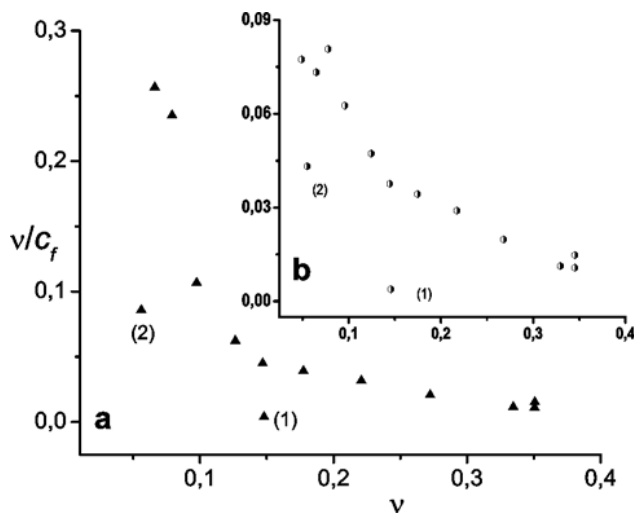}
\caption{The same data as in Fig.~4, but in the Scatchard
coordinates }
\end{figure}

When transforming the experimental data to the Scatchard coordinates (5),
there arises a problem concerning the determination of the bound and free ligand
concentrations: which value should be taken for the optical
parameter for bound ligands? First, let us use the maximal value (188.9) of
optical parameter among all experimental points; the result is shown in
Fig.~8,$a$. It is evident that the spread of points considerably increased
after the linearization.

If the optical parameter for bound ligands is taken from the
approximation results (Fig.~4), we obtain a value of 191.8, which is
rather close to the used one. The corresponding Scatchard plot is
depicted in Fig.~8,$b$. One can see that the plot changed
substantially. Namely, the variation of $A_{b}$ by less than 1.5\%
gave rise to the multiple change of the ratio $\nu/c_{f}$. Moreover,
two points, (1) and (2), drop out of the general tendency. We
excluded them and carried out the direct approximation of the
remained data once more to obtain
$K=(5.64\pm0.35)\times10^{4}~\mathrm{M}^{-1}$ and $n=1.86\pm0.07$.
Both parameters evidently changed considerably. The determination
accuracy for the parameter increased at that, i.e. the omitted
points really inserted a substantial error. We also obtained a
corrected value, $A_{b}=194.3$. It is already clear that the
Scatchard plot essentially depends on this quantity. Therefore, we
replotted it again and found the binding parameters (Fig.~9)
$K=(5.63\pm0.25)\times10^{4}~\mathrm{M}^{-1}$ and $n=1.86\pm0.09$.

\begin{figure}
\includegraphics[width=8.3cm]{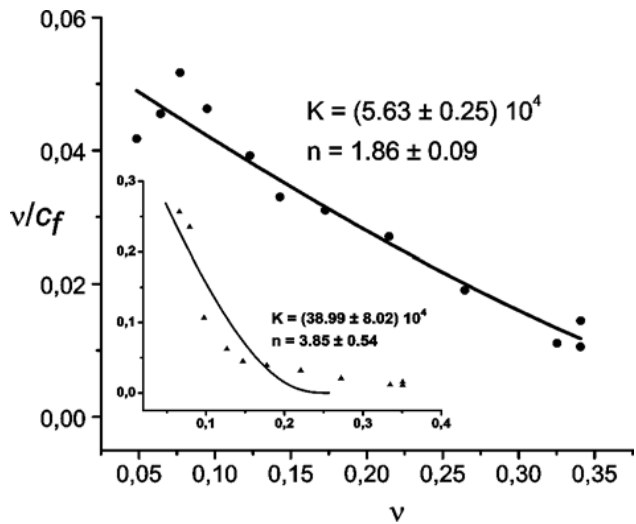}
\caption{Approximation for the non-cooperative binding by the
McGhee--von Hippel equation and in the Scatchard coordinates. The
same, but in the case where $A_{b}$ was determined without the
approximation of initial data, is depicted in the inset  }
\end{figure}

For the sake of comparison, let us approximate the initial
Scatchard plot (Fig.~8,$a$) with omitted points (1) and (2), of
course (see the inset in Fig.~9). Hence, the value obtained for
$A_{b}$ as a result of the direct approximation of experimental
data provides a better agreement between the theoretical curve and
the experimental points plotted in the Scatchard variables
(Fig.~9) than the value obtained without this approximation (the
inset in Fig.~9). The magnitude of association constant strongly
depends on the determination accuracy for $A_{b}$. In our case
where $A_{b}$ changed from 188.9 to 194.3, i.e. the relative
variation of this quantity was less than 3\%, which was close to
its determination error, the association constant changed by an
order of magnitude (see the parameter values in Fig.~9). The error
rescaled into the Scatchard coordinates does not take the
above-mentioned fact into account and, therefore, has not any
reason.

If the transition to the Scatchard coordinates is fulfilled with a sufficient
accuracy, the results practically coincide with those obtained at the direct
approximation. However, this \textquotedblleft sufficient\textquotedblright%
\ accuracy demands for rather a laborious treatment of initial data.

An impression might arise (Fig.~8) that the Scatchard coordinates allow
\textquotedblleft bad\textquotedblright\ experimental points to be excluded
easily. However, at the direct approximation, there are no difficulties in
detecting the points that insert a suspiciously large error. In addition, an advantage
of the program is its equally simple operation with data producing both linear
and nonlinear Scatchard plots.

\section{Conclusions}

Hence, the analysis of the absorption spectra and the fluorescence manifestations of
the interaction between alkaloids berberine and sanguinarine, on the one hand,
and DNA, on the other hand, testifies to the binding of alkaloids with DNA. The
binding character was shown to depend on the concentration ratio between
alkaloid molecules and DNA base pairs. The parameters of the binding of
berberine with DNA were determined with the help of a modified McGhee--von Hippel equation
for the cooperative binding. The obtained values evidence the intercalation
mechanism of binding. We also determined the parameters of
the binding of sanguinarine with DNA. Two types of binding turned out to be characteristic of
the sanguinarine--DNA interaction: the external binding prevails at $N/c<2$ and
the intercalation at $N/c>6$.

\rezume{%
ПАРАМЕТРИ ЗВ'ЯЗУВАННЯ АЛКАЛОЇДІВ БЕРБЕРИНУ ТА САНГВІНАРИНУ З
ДНК}{В.Г. Гуменюк, Н.В. Башмакова, С.Ю.~Кутовий, В.М.~Ящук,
Л.А.~Заїка} {Досліджено взаємодію рослинних алкалоїдів  берберину та
сангвінарину з ДНК у водному розчині методами оптичної спектроскопії
(поглинання, флюоресценція). Розглянуто залежності спектральних
характеристик алкалоїдів від співвідношення концентрацій пар основ
ДНК та молекул алкалоїду ($N/c$), визначено прояви зв'язування
алкалоїдів з ДНК. Показано, що характер зв'язування залежить від
$N/c$. За допомогою модифікованих рівнянь Скетчарда та МакГі--фон
Хіппеля визначено параметри зв'язування  берберину та сангвінарину з
ДНК.}

\end{document}